\documentclass[9pt,twocolumn,twoside]{pnas-new}
\usepackage{color}
\templatetype{pnasresearcharticle} 

\title{From predictions to prescriptions: A data-driven response to COVID-19}

\author[a,b,1]{Dimitris Bertsimas}
\author[b]{Leonard Boussioux} 
\author[b]{Ryan Cory-Wright}
\author[b]{Arthur Delarue}
\author[b]{Vassilis Digalakis}
\author[a,b]{Alexandre Jacquillat}
\author[b]{Driss Lahlou Kitane}
\author[b]{Galit Lukin}
\author[b]{Michael Li}
\author[b]{Luca Mingardi}
\author[c]{Omid Nohadani}
\author[b]{Agni Orfanoudaki}
\author[b]{Theodore Papalexopoulos}
\author[b]{Ivan Paskov}
\author[b]{Jean Pauphilet}
\author[b]{Omar Skali Lami}
\author[b]{Bartolomeo Stellato}
\author[b]{Hamza Tazi Bouardi}
\author[b]{Kimberly Villalobos Carballo}
\author[b]{Holly Wiberg}
\author[b]{Cynthia Zeng}

\affil[a]{Sloan School of Management, Massachusetts Institute of Technology, Cambridge, MA 02142}
\affil[b]{Operations Research Center, Massachusetts Institute of Technology, Cambridge, MA 02139}
\affil[c]{Benefits Science Technologies, Boston, MA 02110}

\leadauthor{Dimitris Bertsimas\hspace{-3 pt}}

\significancestatement{In the midst of the COVID-19 pandemic, healthcare providers and policy makers are wrestling with unprecedented challenges. How to treat COVID-19 patients with equipment shortages? How to allocate resources to combat the disease? How to plan for the next stages of the pandemic? We present a data-driven approach to tackle these challenges. We gather comprehensive data from various sources, including clinical studies, electronic medical records, and census reports. We develop algorithms to understand the disease, predict its mortality, forecast its spread, inform social distancing policies, and re-distribute critical equipment. These algorithms provide decision support tools that have been deployed on our publicly available website, and are actively used by hospitals, companies, and policy makers around the globe.}

\authorcontributions{
D.B., R.C.W., A.D., A.J., D.L.K., M.L., O.N., A.O., I.P., J.P., O.S.L., B.S., H.T.B. and H.W. designed research;
L.B., R.C.W., A.D., V.D., A.J., D.L.K., G.L., M.L., L.M., A.O., T.P., I.P., J.P., O.S.L., B.S., H.T.B., K.V.C., H.W. and C.Z. performed research;
R.C.W., A.D., D.L.K., M.L., L.M., A.O., T.P., I.P., J.P., O.S.L., B.S., H.T.B., and H.W. analyzed data;
D.B., R.C.W., A.D., A.J., M.L., O.N., A.O., J.P., H.T.B. and H.W. wrote the paper.
}
\authordeclaration{No author has any competing interest to declare.}
\correspondingauthor{\textsuperscript{1}To whom correspondence should be addressed. E-mail: dbertsim@mit.edu}


\begin{abstract}
The COVID-19 pandemic has created unprecedented challenges worldwide. Strained healthcare providers make difficult decisions on patient triage, treatment and care management on a daily basis. Policy makers have imposed social distancing measures to slow the disease, at a steep economic price. We design analytical tools to support these decisions and combat the pandemic. Specifically, we propose a comprehensive data-driven approach to understand the clinical characteristics of COVID-19, predict its mortality, forecast its evolution, and ultimately alleviate its impact. By leveraging cohort-level clinical data, patient-level hospital data, and census-level epidemiological data, we develop an integrated four-step approach, combining descriptive, predictive and prescriptive analytics. First, we aggregate hundreds of clinical studies into the most comprehensive database on COVID-19 to paint a new macroscopic picture of the disease. Second, we build personalized calculators to predict the risk of infection and mortality as a function of demographics, symptoms, comorbidities, and lab values. Third, we develop a novel epidemiological model to project the pandemic's spread and inform social distancing policies. Fourth, we propose an optimization model to re-allocate ventilators and alleviate shortages. Our results have been used at the clinical level by several hospitals to triage patients, guide care management, plan ICU capacity, and re-distribute ventilators. At the policy level, they are currently supporting safe back-to-work policies at a major institution and equitable vaccine distribution planning at a major pharmaceutical company, and have been integrated into the US Center for Disease Control’s pandemic forecast.
\end{abstract}

\dates{This manuscript was compiled on \today}
\doi{\url{www.pnas.org/cgi/doi/10.1073/pnas.XXXXXXXXXX}}

\begin{document}

\maketitle
\thispagestyle{firststyle}
\ifthenelse{\boolean{shortarticle}}{\ifthenelse{\boolean{singlecolumn}}{\abscontentformatted}{\abscontent}}{}

\dropcap{I}n just a few weeks, the whole world has been upended by the outbreak of COVID-19, an acute respiratory disease caused by a new coronavirus called SARS-CoV-2. The virus is highly contagious: it is easily transmitted from person to person via respiratory droplet nuclei and can persist on surfaces for days \cite{kampf2020persistence,sanchel2020high}. As a result, COVID-19 has spread rapidly---classified by the World Health Organization as a public health emergency on January 30, 2020 and as a pandemic on March 11. As of mid-May, over 4.5 million cases and 300,000 deaths have been reported globally \cite{JHU}.

Because no treatment is currently available, healthcare providers and policy makers are wrestling with unprecedented challenges. Hospitals and other care facilities are facing shortages of beds, ventilators and personal protective equipment---raising hard questions on how to treat COVID-19 patients with scarce supplies and how to allocate resources to prevent further shortages. At the policy level, most countries have imposed ``social distancing'' measures to slow the spread of the pandemic. These measures allow strained healthcare systems to cope with the disease by ``flattening the curve'' \cite{anderson2020will} but also come at a steep economic price \cite{fernandes2020economic,mckibbin2020global}. Nearly all governments are now confronted to difficult decisions balancing public health and socio-economic outcomes.

This paper proposes a comprehensive data-driven approach to understand the clinical characteristics of COVID-19, predict its mortality, forecast its evolution, and ultimately alleviate its impact. We leverage a broad range of data sources, which include (i) our own cohort-level data aggregating hundreds of clinical studies, (ii) patient-level data obtained from electronic health records, and (iii) census reports on the scale of the pandemic. We develop an integrated approach spanning descriptive analytics (to derive a macroscopic understanding of the disease), predictive analytics (to forecast the near-term impact and longer-term dynamics of the pandemic), and prescriptive analytics (to support healthcare and policy decision-making).

\begin{figure*}[h!]
    \centering
    \includegraphics[width=\linewidth]{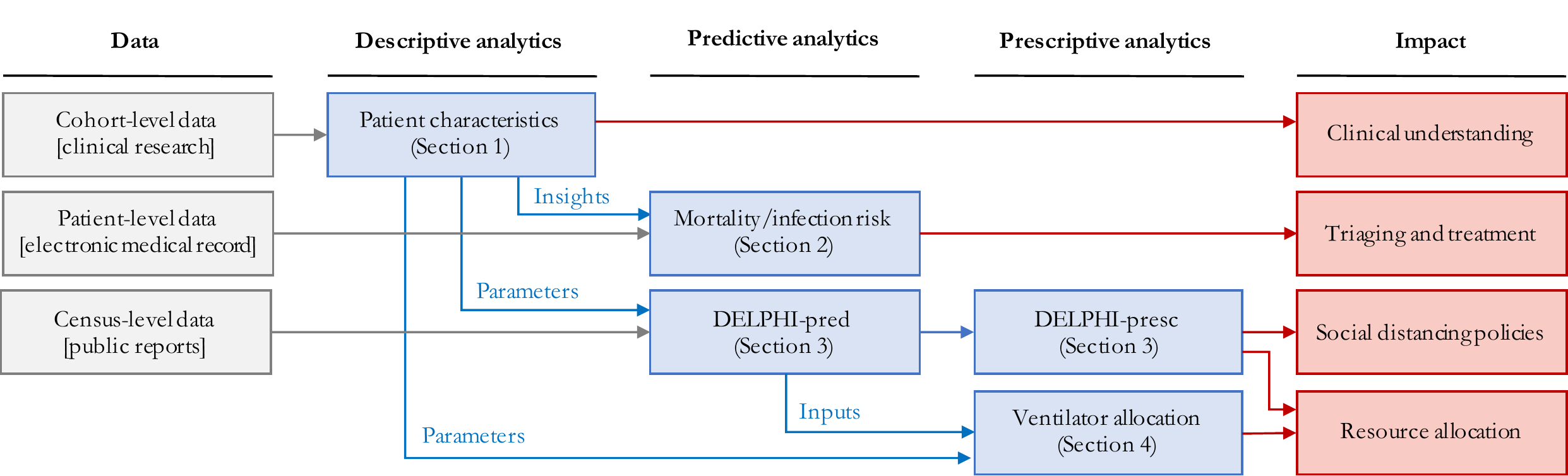}
    \caption{Overview of our end-to-end analytics approach. We leverage diverse data sources to inform a family of descriptive, predictive and prescriptive tools for clinical and policy decision-making support.}
    \label{fig:overview}
\end{figure*}

Specifically, our approach comprises four steps (Figure~\ref{fig:overview}):
\begin{itemize}[topsep=0pt,itemsep=-4pt]
\item	\emph{Aggregating and visualizing the most comprehensive clinical database on COVID-19 (Section 1).} We aggregate cohort-level data on demographics, comorbidities, symptoms and lab values from 160 clinical studies. These data paint a broad picture of the disease, identifying common symptoms, disparities between mild and severe patients, and geographic disparities---insights that are hard to derive from any single study and can orient future clinical research on COVID-19, its mutations, and its disparate effects across ethnic groups.
\item	\emph{Providing personalized indicators to assess the risk of mortality and infection (Section 2).} Using patient-level data, we develop machine learning models to predict mortality and infection risk, as a function of demographics, symptoms, comorbidities, and lab values. Using gradient boosting methods, the models achieve strong predictive performance---with an out-of-sample area under the curve above 90\%. These models yield personalized calculators that can (i) guide triage, treatment, and care management decisions for strained healthcare systems, and (ii) serve as pre-screening tools for patients before they visit healthcare or testing facilities.
\item	\emph{Developing a novel epidemiological model to forecast the evolution of the disease and assess the effects of social distancing (Section 3).} We propose a new compartmental model called DELPHI, which accounts for COVID-19 features such as underdetection and government response. The model estimates the disease's spread with high accuracy; notably, its projections from as early as April 3 have matched the number of cases observed in the United States up to mid-May. We also provide a data-driven assessment of social distancing policies, showing that the pandemic's spread is highly sensitive to the stringency and timing of mitigating measures.
\item	\emph{Proposing an optimization model to support ventilator allocation in response to the pandemic (Section 4).} We formulate a mixed-integer optimization model to allocate ventilators efficiently in a semi-collaborative setting where resources can be shared both between healthcare facilities or through a central authority. In the United States, this allows us to study the trade-offs of managing the federal ventilator stockpile in conjunction with inter-state transfers. Results show that limited ventilator transfers could have eliminated shortages in April 2020.
\end{itemize}

A major contribution of our work is to treat these different questions as interdependent challenges raised by the pandemic---as opposed to a series of isolated problems. Indeed, clinical decision-making depends directly on patient inflows and available supplies, while resource planning and government responses react to patient-level outcomes. By combining various data sources into descriptive, predictive and prescriptive methods, this paper proposes an end-to-end approach to design a comprehensive and cohesive response to COVID-19.

Ultimately, this paper develops analytical tools to inform clinical and policy responses to the COVID-19 pandemic. These tools are available to the public on a dedicated website.\footnote{\href{https://www.covidanalytics.io}{\texttt{www.covidanalytics.io}}} They have also been deployed in practice to combat the spread of COVID-19 globally. Several hospitals in Europe have used our risk calculators to support pre-triage and post-triage decisions, and a major financial institution in South America is applying our infection risk calculator to determine how employees can safely return to work. A major hospital system in the United States planned its intensive care unit (ICU) capacity based on our forecasts, and leveraged our optimization results to allocate ventilators across hospitals when the number of cases was rising. Our epidemiological predictions are used by a major pharmaceutical company to design a vaccine distribution strategy that can contain future phases of the pandemic. They have also been incorporated into the US Center for Disease Control's forecasts \cite{CDC}.

\section{Descriptive Analytics: Clinical Outcomes Database} \label{sec:outcomes_database}
    
Early responses to the COVID-19 pandemic have been inhibited by the lack of available data on patient outcomes. Individual centers released reports summarizing patient characteristics. Yet, this decentralized effort makes it difficult to construct a cohesive picture of the pandemic.

To address this problem, we construct a database that aggregates demographics, comorbidities, symptoms, laboratory blood test results (``lab values'', henceforth) and clinical outcomes from $160$ clinical studies released between December $2019$ and May $2020$---made available on our website for broader use. The database contains information on 133,600 COVID-$19$ patients (3.13\% of the global COVID-19 patients as of May 12, 2020), spanning mainly Europe ($81,207$ patients), Asia ($19,418$ patients) and North America ($23,279$ patients). To our knowledge, this is the largest dataset on COVID-19.

\subsection{Data Aggregation} Each study was read by an MIT researcher, who transcribed numerical data from the manuscript. The appendix reports the main transcription assumptions.

Each row in the database corresponds to a cohort of patients---some papers study a single cohort, whereas others study several cohorts or sub-cohorts. Each column reports cohort-level statistics on demographics (e.g., average age, gender breakdown), comorbidities (e.g., prevalence of diabetes, hypertension), symptoms (e.g.,  prevalence of fever, cough),  treatments (e.g., prevalence of antibiotics, intubation), lab values (e.g., average lymphocyte count), and clinical outcomes (e.g., average hospital length of stay, mortality rate). We also track whether the cohort comprises ``mild'' or ``severe'' patients (mild and severe cohorts are only a subset of the data).

Due to the pandemic's urgency, many papers were published before all patients in a cohort were discharged or deceased. Accordingly, we estimate the mortality rate from discharged and deceased patients only (referred to as ``Projected Mortality'').

\subsection{Objectives}
Our main goal is to leverage this database to derive a macroscopic understanding of the disease. We break it down into the following questions:
\begin{itemize}[topsep=0pt,itemsep=-4pt]
    \item Which symptoms are most prevalent?
    \item How do ``mild'' and ``severe'' patients differ in terms of symptoms, comorbidities, and lab values?
    \item Can we identify epidemiological differences in different parts of the world?
\end{itemize}

\subsection{Descriptive Statistics}

Table \ref{ref:tabsymptprev} depicts the prevalence of COVID-19 symptoms, in aggregate, classified into ``mild'' or ``severe'' patients, and classified per geographic region. Our key observations are that:
\begin{itemize}[topsep=0pt,itemsep=-4pt]
    \item Cough, fever, shortness of breath, and fatigue are the most prevalent symptoms of COVID-19.
    \item COVID-19 symptoms are much more diverse than those listed by public health agencies. COVID-19 patients can experience at least 15 different symptoms. In contrast, the US Center for Disease Control and Prevention lists seven symptoms (cough, shortness of breath, fever, chills, myalgia, sore throat, and loss of taste/smell) \cite{cdc2020}; the World Health Organization lists three symptoms (fever, cough, and fatigue) \cite{who2020}; and the UK National Health Service lists two main symptoms (fever and cough) \cite{nhs2020}. This suggests a lack of consensus among the medical community, and opportunities to revisit public health guidelines to capture the breadth of observed symptoms.
    \item Shortness of breath and elevated respiratory rates are much more prevalent in cases diagnosed as severe.
    \item Symptoms are quite different in Asia vs. Europe or North America. In particular, more than 75\% of Asian patients experience fever, as compared to less than half in Europe and North America. Conversely, shortness of breath is much more prevalent in Europe and North America.
\end{itemize}

\begin{table*}[h!]\centering
\caption{Count and prevalence of symptoms among COVID-$\boldsymbol{19}$ patients, in aggregate, broken down into mild/severe patients, and broken down per continent (Asia, Europe, North America). Mild and severe patients only form a subset of the data, and so do patients from Asia, Europe and North America. A ``-'' indicates that fewer than $\boldsymbol{100}$ patients in a subpopulation reported on this symptom.}
\begin{tabular}{l rr r rr r rr r rr r rr r rr}\toprule 
     Symptom & \multicolumn{2}{c}{All patients} & & \multicolumn{2}{c}{Mild} & & \multicolumn{2}{c}{Severe} & & \multicolumn{2}{c}{Asia} & & \multicolumn{2}{c}{Europe} & & \multicolumn{2}{c}{North America} \\
    \cmidrule{2-3} \cmidrule{5-6} \cmidrule{8-9} \cmidrule{11-12} \cmidrule{14-15} \cmidrule{17-18}
     & Count & $(\%)$ &
     & Count & $(\%)$ &
     & Count & $(\%)$ &
     & Count & $(\%)$ &
     & Count & $(\%)$ &
     & Count & $(\%)$\\\midrule 
    Cough & $94,950$ & $52.8\%$ & & $6,833$ & $63.0\%$ & & $5,803$ & $50.4\%$ && $14,034$ & $56.2\%$ & & $78,430$ & $52.2\%$ & & $1,113$ & $63.6\%$\\
    Fever & $95,870$ & $48.1\%$ & & $6,864$ & $79.3\%$ & & $6,077$ & $76.7\%$ && $14,750$ & $76.6\%$ & & $78,450$ & $43.5\%$ & & $1,481$ & $41.3\%$\\
    Short Breath & $17,290$ & $33.7\%$ & & $6,006$ & $16.1\%$ & & $5,373$ & $60.7\%$ && $11,330$ & $19.7\%$ & & $3,512$ & $69.9\%$ & & $1,111$ & $49.2\%$\\
    Fatigue & $11,560$ & $31.4\%$ & & $5,313$ & $35.3\%$ & & $1,989$ & $40.6\%$ && $11,320$ & $30.8\%$ & & $226$ & $64.2\%$ & & $-$ & $-$\\
    Sputum & $7,613$ & $26.3\%$ & & $4,995$ & $29.2\%$ & & $1,216$ & $34.2\%$ && $7,395$ & $26.7\%$ & & $-$ & $-$ & & $176$ & $10.9\%$\\
    Sore Throat & $83,170$ & $22.2\%$ & & $3,513$ & $14.2\%$ & & $921$ & $8.2\%$ && $6,013$ & $10.4\%$ & & $75,235$ & $22.9\%$ & & $550$ & $9.8\%$\\
    Myalgia & $12,150$ & $17.5\%$ & & $4,455$ & $16.4\%$ & & $1,643$ & $19.1\%$ && $8,517$ & $15.5\%$ & & $1,633$ & $33.5\%$ & & $755$ & $25.3\%$\\ 
    Elev. Resp. Rate & $7,376$ & $16.4\%$ & & $527$ & $9.7\%$ & & $642$ & $38.4\%$ && $1,257$ & $14.6\%$ & & $-$ & $-$ & & $6,117$ & $16.8\%$\\
    Anorexia & $3,928$ & $15.8\%$ & & $1,641$ & $14.2\%$ & & $808$ & $15.4\%$ && $3,566$ & $13.8\%$ & & $312$ & $40.5\%$ & & $-$ & $-$\\
    Headache & $11,430$ & $15.7\%$ & & $5,068$ & $12.2\%$ & & $1,541$ & $8.6\%$ && $7,929$ & $9.9\%$ & & $1,633$ & $27.2\%$ & & $551$ & $8.7\%$\\
    Nausea & $10,070$ & $12.4\%$ & & $4,238$ & $6.5\%$ & & $1,798$ & $5.6\%$ && $8,262$ & $8.2\%$ & & $312$ & $22.4\%$ & & $259$ & $9.0\%$\\
    Chest Pain & $3,303$ &$11.3\%$ & & $767$ & $12.2\%$ & & $588$ & $19.6\%$ && $2,984$ & $12.2\%$ & & $-$ & $-$ & & $-$ & $-$\\
    Diarrhea & $16,520$ & $11.1\%$ & & $5,687$ & $9.7\%$ & & $5,369$ & $9.0\%$ && $11,470$ & $10.8\%$ & & $3,512$ & $10.4\%$ & & $1,066$ & $15.4\%$\\
    Cong. Airway & $1,639$ & $8.7\%$ & & $2,176$ & $6.5\%$ & & $234$ & $14.1\%$ && $1,369$ & $8.9\%$ & & $-$ & $-$ & & $258$ & $7.4\%$\\
    Chills & $3,116$ & $8.7\%$ & & $2,751$ & $9.9\%$ & & $520$ & $9.4\%$ && $2,794$ & $8.2\%$ & & $-$ & $-$ & & $268$ & $11.5\%$\\
    \midrule
    Proj. Mortality & $111,700$ & $11.7\%$ && $7,428$ & $0.4\%$ && $9,146$ & $74.0\%$ && $12,820$ & $16.7\%$ & & $79,750$ & $9.9\%$ & & $19,060$ & $15.8\%$  \\
     \bottomrule
\end{tabular}
\label{ref:tabsymptprev}
\end{table*}

\begin{figure*}[h!]
    \centering
    \includegraphics[width=1.7\columnwidth]{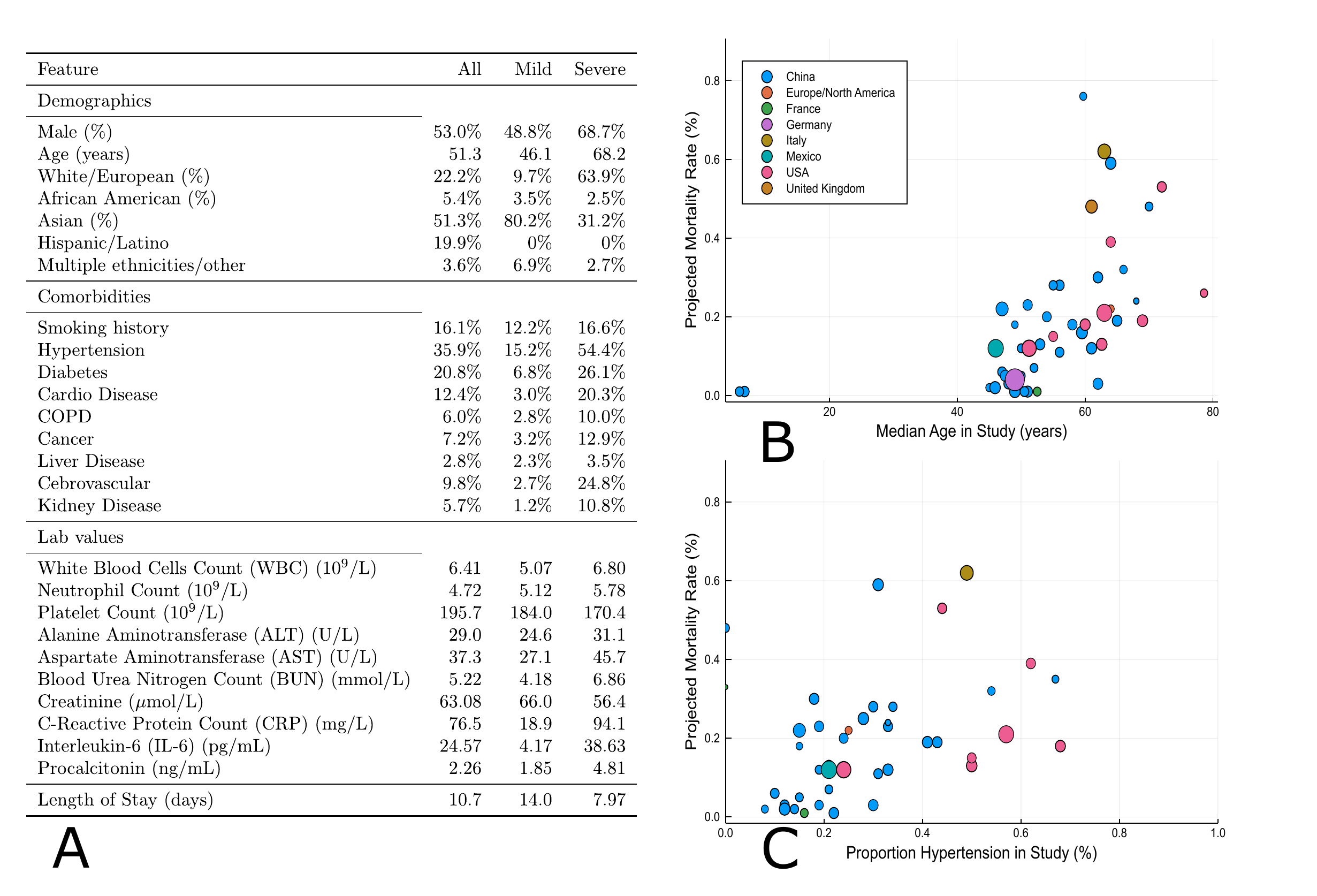}
    \caption{Summary of demographics, comorbidities and lab values in mild and severe COVID-19 patients. (A) Comorbidities, demographics, average lab values, average length of stay and projected mortality among COVID-$19$ patients, in aggregate and broken down into mild/severe patients. (B) Impact of median age on projected mortality at a cohort level. (C) Impact of hypertension rates on projected mortality at a cohort level. The size of each dot represents the number of patients in the cohort, and its color represents the nation the study was performed in. We only include studies reporting both discharged and deceased patients.}
    \label{fig:demos_comborbs_vs_mort}
\end{figure*}

Using a similar nomenclature, Figure \ref{fig:demos_comborbs_vs_mort}A reports demographics, comorbidities, lab values, and clinical outcomes (an extended version is available in the appendix). In terms of demographics, severe populations of patients have a higher incidence of male subjects and are older on average. Severe patients also have elevated comorbidity rates. Figures \ref{fig:demos_comborbs_vs_mort}B and~\ref{fig:demos_comborbs_vs_mort}C visually confirm the impact of age and hypertension rates on population-level mortality---consistent with \cite{guan2020clinical,goyal2020clinical, petrilli2020factors}. In terms of lab values, CRP, AST, BUN, IL-6 and Protocalcitonin are highly elevated among severe patients.

\subsection{Discussion and Impact}
Our database is the largest available source of clinical information on COVID-19 assembled to date. As such, it provides new insights on common symptoms and the drivers of the disease's severity. Ultimately, this database can support guidelines from health organizations, and contribute to ongoing clinical research on the disease.

Another benefit of this database is its geographical reach. Results highlight disparities in patients' symptoms across regions. These disparities may stem from (i) different reporting criteria; (ii) different treatments; (iii) disparate impacts across different ethnic groups; and (iv) mutations of the virus since it first appeared in China. This information contributes to early evidence on COVID-19 mutations \citep{forster2020phylogenetic, holland202081} and on its disparate effects on different ethnic groups \citep{garg2020hospitalization,vahidy2020racial}.

Finally, the database provides average values of key parameters into our epidemiological model of the disease's spread and our optimization model of resource allocation (e.g., average length of stay of hospitalizations, average fraction of hospitalized patients put on a ventilator).

The insights derived from this descriptive analysis highlight the need for personalized data-driven clinical indicators. Yet, our population-level database cannot be leveraged directly to support decision-making at the patient level. We have therefore initiated a multi-institution collaboration to collect electronic medical records from COVID-19 patients and develop clinical risk calculators. These calculators, presented in the next section, are informed by several of our descriptive insights. Notably, the disparities between severe patients and the rest of the patient population inform the choice of the features included in our mortality risk calculator. Moreover, the geographic disparities suggest that data from Asia may be less predictive when building infection or mortality risk calculators designed for patients in Europe or North America---motivating our use of data from Europe.

\section{Predictive Analytics: Mortality and Infection Risk}\label{sec:calculators}

Throughout the COVID-19 crisis, physicians have made difficult triage and care management decisions on a daily basis. Oftentimes, these decisions could only rely on small-scale clinical tests, each requiring significant time, personnel and equipment and thus cannot be easily replicated. As the burden on ``hot spots” has ebbed, hospitals began to aggregate rich data on COVID-19 patients. This data offers opportunities to develop algorithmic risk calculators for large-scale decision support---ultimately facilitating a more proactive and data-driven strategy to combat the disease globally.

We have established a patient-level database of thousands of COVID-19 hospital admissions. Using state-of-the-art machine learning methods, we develop a \emph{mortality risk calculator} and an \emph{infection risk calculator}. Together, these two risk assessments provide screening tools to support critical care management decisions, spanning patient triage, hospital admissions, bed assignment and testing prioritization. A more detailed model for mortality  with lab values is presented in 
\cite{db_mortality}.

\subsection{Methods}

This investigation constitutes a multi-center study from healthcare institutions in Spain and Italy, two countries severely impacted by COVID-19. Specifically, we collected data from (i) Azienda Socio-Sanitaria Territoriale di Cremona (ASST Cremona), the main hospital network in the Province of Cremona, and (ii) HM Hospitals, a leading hospital group in Spain with 15 general hospitals and 21 clinical centers spanning the regions of Madrid, Galicia, and León. We applied the following inclusion criteria to the calculators:
\begin{itemize}
    \item \textbf{Mortality Risk}: We include adult patients diagnosed with COVID-19 and hospitalized. We consider patients who were either discharged from the hospital or deceased within the visit---excluding active patients. We include only lab values and vital values collected on the first day in the emergency department to match the clinical decision setting---predicting prognosis at the time of admission.
   \item \textbf{Infection Risk}: We include adult patients who underwent a polymerase chain reaction test for detecting COVID-19 infection at the ASST Cremona hospital~\cite{lan2020positive}.\footnote{HM Hospitals patients were not included since no negative case data was available.} We include all patients, regardless of their clinical outcome. Each patient was subject to a blood test. We omit comorbidities since they are derived from the discharge diagnoses, hence not available for all patients. 
\end{itemize}


We train two models for each calculator: one with lab values and one without lab values. Missing values are imputed using $k$-nearest neighbors imputation~\cite{knnImputations}. We exclude features missing for more than 40\% of patients. We train binary classification models for both risk calculators, using the XGBoost algorithm~\cite{chen2016xgboost}. We restrict the model to select at most 20 features, in order to make the resulting tool easily usable. We use SHapley Additive exPlanations (SHAP)~\cite{NIPS2017_7062,lundberg2020local2global} to generate importance plots that identify risk drivers and provide transparency on the model predictions.

To evaluate predictive performance, we use 40 random data partitions into training and test sets. We compute the average Area Under the Curve (AUC), sensitivity, specificity, precision, negative predictive value, and positive predictive value. We calculate 95\% confidence intervals using bootstrapping.

\subsection{Results}

\subsubsection*{Study Population}
The mortality study population comprises 2,831 patients, 711 (25.1\%) of whom died during hospitalization while the remaining ones were discharged. The infection study population comprises 3,135 patients, 1,661 (53.0\%) of whom tested positive for COVID-19. The full distributions of patient characteristics are reported in the appendix.

\subsubsection*{Performance Evaluation}
All models achieve strong out-of-sample performance. Our mortality risk calculator has an AUC of 93.8\% with lab values and 90.5\% without lab values. Our infection risk calculator has an AUC of 91.8\% with lab values and 83.1\% without lab values. These values suggest a strong discriminative ability of the proposed models. We report in the appendix average results across all random data partitions.

We also report in the appendix threshold-based metrics, which evaluate the discriminative ability of the calculators at a fixed cutoff. With the threshold set to ensure a sensitivity of at least 90\% (motivated by the high costs of false negatives), we obtain accuracies spanning 65\%--80\%.

The mortality model achieves better overall predictive performance than the infection model. As expected, both models have better predictive performance with lab values than without lab values. Yet, the models without lab values still achieve strong predictive performance.

\subsubsection*{Model Interpretation}
\begin{figure*}[h!]
\centering
\includegraphics[width=2\columnwidth]{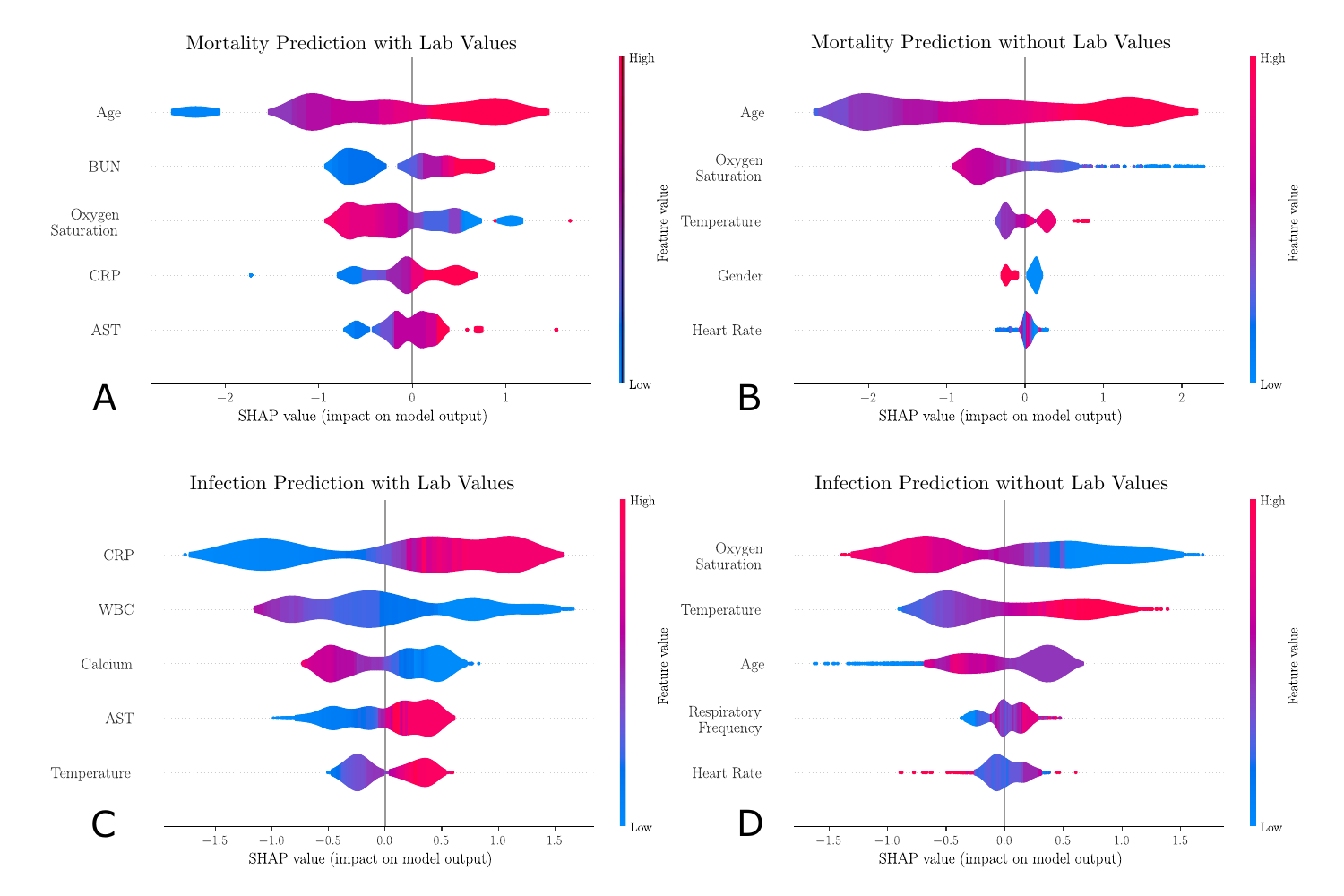}
\caption{SHapley Additive exPlanations (SHAP) importance plots for the mortality and infection risk calculators, including: (A) the mortality model with lab values; (B) the mortality model without lab values; (C) the infection model with lab values; and (D) the infection model without lab values. The five most important features are shown for each model. Gender is a binary feature (female is equal to 1, shown in red; male is equal to 0, shown in blue). Each row represents the impact of a feature on the outcome, with higher SHAP values indicating higher likelihood of a positive outcome.}
\label{fig:shap}
\end{figure*}

Figure~\ref{fig:shap} plots the SHAP importance plots for all models. The figures sort the features by decreasing significance. For each one, the row represents its impact on the SHAP value, as the feature ranges from low (blue) to high (red). Higher SHAP values correspond to increased likelihood of a positive outcome (i.e. mortality or infection). Features with the color scale oriented blue to red (resp. red to blue) from left to right have increasing (resp. decreasing) risk as the feature increases. For example, ``Age'' is the most important feature of the mortality score with lab values (Figure~\ref{fig:shap}A), and older patients have higher predicted mortality.

\subsection{Discussion and Impact}

The models with lab values provide algorithmic screening tools that can deliver COVID-19 risk predictions using common clinical features. In a constrained healthcare system or in a clinic without access to advanced diagnostics, clinicians can use these models to rapidly identify high-risk patients to support triage and treatment decisions.

The models without lab values offer an even simpler tool that could be used outside of a clinical setting. In strained healthcare systems, it can be difficult for patients to obtain direct advice from providers. Our tool could serve as a pre-screening step to identify personalized infection risk---without visiting a testing facility. While the exclusion of lab values reduces the AUC (especially for infection), these calculators still perform strongly.

Our models provide insights into risk factors and biomarkers related to COVID-19 infection and mortality. Our results suggest that the main indicators of mortality risk are age, BUN, CRP, AST, and low oxygen saturation. These findings validate several population-level insights from Section~\ref{sec:outcomes_database} and are in agreement with clinical studies: prevalence of shortness of breath~\cite{wang2020unique}, elevated levels of CRP as an inflammatory marker~\cite{velavan2020covid,caruso2020chest}, and elevated AST levels due to liver dysfunction in severe COVID-19 cases~\cite{guan2020clinical,huang2020clinical}. 

Turning to infection risk, the main indicators are CRP, WBC, Calcium, AST, and temperature. These findings are also in agreement with clinical reports: an elevated CRP generally indicates an early sign of infection and implies lung lesions from COVID-19~\cite{ling2020c}, elevated levels of leukocytes suggest cytokine release syndrome caused by SARS-CoV-2 virus~\cite{shi2020covid}, and lowered levels of serum calcium signal higher rate of organ injury and septic shock~\cite{shi2020serum}. The agreement between our findings and clinicl observations offers credibility for the use of our calculators to support clinical decision-making---although they are not intended to substitute clinical diagnostic or medical expertise.


When lab values are not available, the widely accepted risk factors of age, oxygen saturation, temperature, and heart rate become the key indicators for both risk calculators. We observe that mortality risk is higher for male patients (blue in  Figure~\ref{fig:shap}B) than for female patients (red), confirming clinical reports~\cite{jordan2020covid,liu2020neutrophil}. An elevated respiratory frequency becomes an important predictor of infection, as reported in~\cite{zhou2020clinical}. These findings suggest that demographics and vitals provide valuable information in the absence of lab values. However, when lab values are available, these other features become secondary.

A limitation of the current mortality model is that it does not take into account medication and treatments during hospitalization. We intend to incorporate these in future research to make these models more actionable. Furthermore, these models aim to reveal associations between risks and patient characteristics but are not designed to establish causality.

Overall, we have developed data-driven calculators that allow physicians and patients to assess mortality and infection risks in order to guide care management---especially with scarce healthcare resources. These calculators are being used by several hospitals within the ASST Cremona system to support triage and treatment decisions---alleviating the toll of the pandemic. Our infection calculator also supports safety protocols for Banco de Credito del Peru, the largest bank in Peru, to determine how employees can return to work.

\section{Predictive and Prescriptive Analytics: Disease Projections and Government Response}\label{sec:delphi}

We develop a new epidemiological model, called DELPHI (Differential Equations Leads to Predictions of Hospitalizations and Infections). The model first provides a predictive tool to forecast the number of detected cases, hospitalizations and deaths---we refer to this model as ``DELPHI-pred''. It then provides a prescriptive tool to simulate the effect of policy interventions and guide government response to the COVID-19 pandemic---we refer to this model as ``DELPHI-presc''. All models are fit in each US state (plus the District of Columbia). A detailed presentation and discussion on the implications of the DELPHI model especially with respect to government interventions is presented in \cite{delphi}.

\subsection{DELPHI-pred: Projecting Early Spread of COVID-19}

\subsubsection{Model Development}

DELPHI is a compartmental model, with dynamics governed by ordinary differential equations. It extends the standard SEIR model by defining 11 states (Figure~\ref{fig:delphi_panel}A): susceptible ($S$), exposed ($E$), infectious ($I$), undetected people who will recover ($U_R$) or decease ($U_D$), detected hospitalized people who will recover ($DH_R$) or decease ($DH_D$), quarantined people who will recover ($DQ_R$) or decease ($DQ_D$), recovered ($R$) and deceased ($D$). The separation of the $U_R$/$U_D$, $DQ_R$/$DQ_D$ and $DH_R$/$DH_D$ states enables separate fitting of recoveries and deaths from the data.
    






As opposed to other COVID-19 models \cite[see, e.g.,][]{kissler2020projecting}, DELPHI captures two key elements of the pandemic:
\begin{itemize}[topsep=0pt,itemsep=-4pt]
    \item \textbf{Underdetection}: Many cases remain undetected due to limited testing, record failures, and detection errors. Ignoring them would underestimate the scale of the pandemic. We capture them through the $U_R$ and $U_D$ states.
    \item \textbf{Government Response}: ``Social distancing'' policies limit the spread of the virus. Ignoring them would overestimate the spread of the pandemic. We model them through a decline in the infection rate over time. Specifically, we write:
    $\frac{\mathrm{d} S}{\mathrm{d} t} = -\alpha \gamma(t)S(t)I(t),$
    where $\alpha$ is a constant baseline rate and $\gamma(t)$ is a time-dependent function characterizing each state's policies, modeled as follows:
    \[\gamma(t)=\frac{2}{\pi}\arctan\left(\frac{-(t-t_0)}{k}\right)+1.\]

The inverse tangent function provides a concave-convex relationship, capturing three phases of government response. In \emph{Phase I}, most activities continue normally as people adjust their behavior. In \emph{Phase II}, the infection rate declines sharply as policies are implemented. In \emph{Phase III}, the decline in the infection rate reaches saturation. The parameters $t_0$ and $k$ can be respectively thought of as the start date and the strength of the response.
\end{itemize}

Ultimately, DELPHI involves 13 parameters that define the transition rates between the 11 states. We calibrate six of them from our clinical outcomes database (Section~\ref{sec:outcomes_database}). Using non-linear optimization, we estimate seven parameters for each US state from the data to minimize in-sample error. This training procedure leverages historical data on the number of cases and deaths per US county \cite{NYT}. We include each state as soon as it records more than 100 cases. We provide details on the fitting procedure in the appendix.

\subsubsection{Validation}

\begin{figure*}[h!]
\centering
\includegraphics[width=2.1\columnwidth]{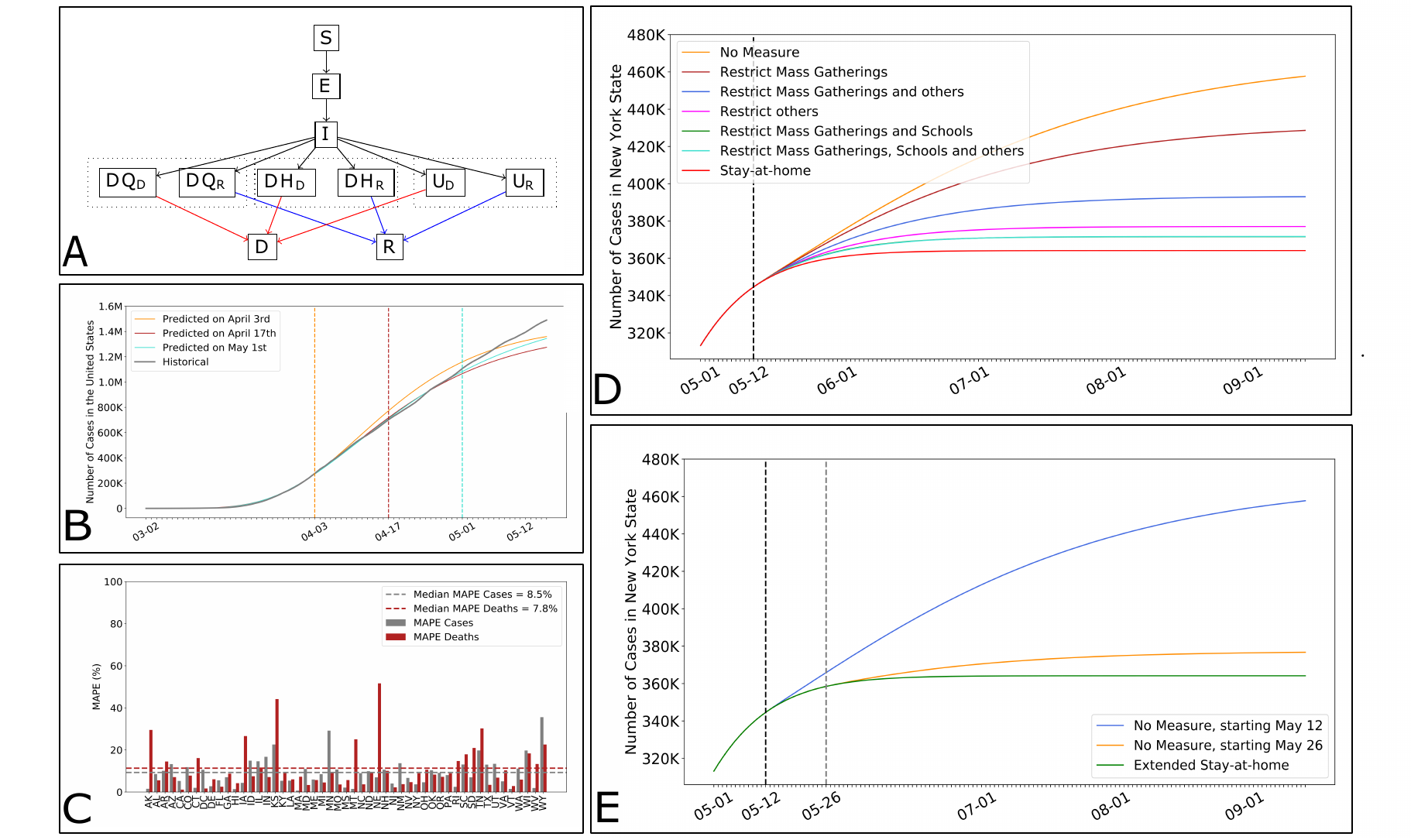}
\caption{DELPHI, an epidemiological model to guide government response. (A) Simplified flow diagram of DELPHI. (B) Cumulative number of cases in the United States according to our projections made at different points in time, against actual observations. (C) Out-of-sample Mean Absolute Percentage Error (MAPE) on the number of cases and deaths per US state. (D) Impact of different policies on the future number of cases, in NY. (E) Impact of the timing of policies on the future number of cases, in NY.}
\label{fig:delphi_panel}
\end{figure*}

DELPHI was created in late March and has been continuously updated to reflect new observed data. Figure~\ref{fig:delphi_panel}B shows our projections made on three different dates, and compares them against historical observations. This plot focuses on the number of cases, but a similar plot for the number of deaths is reported in the appendix.

In addition to providing aggregate validation figures, we also evaluate the model's out-of-sample performance quantitatively, using a backtesting procedure. To our knowledge, this represents the first attempt to assess the predictive performance of COVID-19 projections. Specifically, we fit the model's parameters using data up to April 27, build projections from April 28 to May 12, and evaluate the resulting Mean Absolute Percentage Error (MAPE). Figure~\ref{fig:delphi_panel}C reports the results in each US state.

\subsubsection{Discussion and Impact}

Results suggest that DELPHI-pred achieves strong predictive performance. The model has been consistently predicting, with high accuracy the overall spread of the disease for several weeks. Notably, DELPHI-pred was able to anticipate, as early as April 3rd, the dynamics of the pandemic in the United States up to mid-May. At a time where 200,000--300,000 cases were reported, the model was predicting 1.2M--1.4M cases by mid-May---a prediction that proved accurate 40 days later.

Our quantitative results confirm the visual evidence. The MAPE is small across US states. The median MAPE is 8.5\% for the number of cases---the 10\% and 90\% percentiles are equal to 1.9\% and 16.7\%. The median MAPE is 7.8\% for the number of deaths---the 10\% and 90\% percentiles are equal to 3.3\% and 25.1\%. Given the high level of uncertainty and variability in the disease's spread, this level of accuracy is suggestive of excellent out-of-sample performance.

As Figure~\ref{fig:delphi_panel}C shows, a limitation of our model is that the relative error remains large for a small minority of US states. These discrepancies stem from two main reasons. First, errors are typically larger for states that have recorded few cases (WY) or few deaths (AK, KS, NE). Like all SEIR-derived models, DELPHI performs better on large populations. Moreover, the MAPE metric emphasizes errors on smaller population counts. Second, our model is fitted at the state level, implicitly assuming that the spread of the pandemic is independent from one state to another---thus ignoring inter-state travel. This limitation helps explain the above-median error in a few heartland states which were confronted to the pandemic in later stages (MN, TN, IA).

In summary, DELPHI-pred is a novel epidemiological model of the pandemic, which provides high-quality estimates of the daily number of cases and deaths per US state. This model has been incorporated to the forecasts used by the US Center for Disease Control to chart and anticipate the spread of the pandemic \cite{CDC}. It has also been used by the Hartford HealthCare system---the major hospital system in Connecticut, US---to plan its ICU capacity, and by a major pharmaceutical company to design a vaccine distribution strategy that can most effectively contain the next phases of the pandemic.

\subsection{DELPHI-presc: Toward Re-opening Society}\label{subsec:DELPHI-presc-core}

To inform the relaxation of social distancing policies, we link policies to the infection rate using machine learning. Specifically, we predict the values of $\gamma(t)$, obtained from the fitting procedure of DELPHI-pred. For simplicity and interpretability, we consider a simple model based on regression trees \cite{breiman1984classification} and restrict the independent variables to the policies in place. We classify policies based on whether they restrict mass gatherings, school and/or other activities (referred to as ``Others'', and including business closures, severe travel limitations and/or closing of non-essential services). We define a set of seven mutually exclusive and collectively exhaustive policies observed in the US data: (i) \emph{No measure}; (ii) \emph{Restrict mass gatherings}; (iii) \emph{Restrict others}; (iv) \emph{Authorize schools, restrict mass gatherings and others}; (v) \emph{Restrict mass gatherings and schools}; (vi) \emph{Restrict mass gatherings, schools and others}; and (vii) \emph{Stay-at-home}.

We report the regression tree in the appendix, obtained from state-level data in the United States. This model achieves an out-of-sample $R^2$ of 0.8, suggesting a good fit to the data. As expected, more stringent policies lead to lower values of $\gamma(t)$. The results also provide comparisons between various policies---for instance, school closures seem to induce a stronger reduction in the infection rate than restricting ``other'' activities. More importantly, the model quantifies the impact of each policy on the infection rate. We then use these results to predict the value of $\gamma(t)$ as a function of the policies (see appendix for details), and simulate the spread of the disease as states progressively loosen social distancing policies.

Figure~\ref{fig:delphi_panel}D plots the projected case count in the state of New York (NY), for different policies (we report a similar plot for the death count in the appendix). Note that the stringency of the policies has a significant impact on the pandemic's spread and ultimate toll. For instance, relaxing all social distancing policies on May 12 can increase the \emph{cumulative} number of cases in NY by up to 25\% by September. 

Using a similar nomenclature, Figure~\ref{fig:delphi_panel}E shows the case count if all social distancing policies are relaxed on May 12 vs. May 26. Note that the timing of the policies also has a strong impact: a two-week delay in re-opening society can greatly reduce a resurgence in NY.

The road back to a new normal is not straightforward: results suggest that the disease's spread is highly sensitive to both the intensity and the timing of social distancing policies. As governments grapple with an evolving pandemic, DELPHI-presc can be a useful tool to explore alternative scenarios and ensure that critical decisions are supported with data.


\section{Prescriptive Analytics: Ventilator Allocation}\label{sec:ventilator}


COVID-19 is primarily an acute respiratory disease. The World Health Organization recommends that patients with oxygen saturation levels below $93\%$ receive respiratory support \cite{who2020}. Following the standard Acute Respiratory Distress Syndrome protocol, COVID-19 patients are initially put in the prone position and then put in a drug induced paralysis via a neuromuscular blockade to prevent lung injury~\cite{cornejo2013effects}. Patients are then put on a ventilator, which delivers high concentrations of oxygen while removing carbon dioxide~\cite{bein2016standard}. Early evidence suggests that ventilator intubation reduces the risk of hypoxia for COVID-19 patients~\cite{meng2020intubation}.

As a result, hospitals have been facing ventilator shortages worldwide \cite{ranney2020critical}. Still, local shortages do not necessarily imply global shortages. For instance, in April 2020, the total supply of ventilators in the United States exceeded the projected demand from COVID-19 patients. Ventilator shortages could thus be alleviated by pooling the supply, i.e., by strategically allocating the surge supply of ventilators from the federal government and facilitating inter-state transfers of ventilators.

We propose an optimization model to support the allocation of ventilators in a semi-collaborative setting where resources can be shared both between healthcare facilities or through a central authority. Based on its primary motivation, we formulate the model to support the management of the federal supply of ventilators and inter-state ventilator transfers in the United States. A similar model has also been used to support inter-hospital transfers of ventilators. The model can also support inter-country ventilator allocation during the next phases of the pandemic. This model leverages the demand projections from DELPHI-pred (Section~\ref{sec:delphi}) to prescribe resource allocation recommendations---with the ultimate goal of alleviating the health impact of the pandemic.



\subsection{Model}

Resource allocation is critical when clinical care depends on scarce equipment. Several studies have used optimization to support ventilator pooling. A time-independent model was first developed for influenza planning \cite{huang2017stockpiling}. A time-dependent stochastic optimization model was developed to support transfers to and from the federal government for COVID-19, given scenarios regarding the pandemic's spread \cite{mehrotra2020model}. In this section, we propose a deterministic time-dependent model, leveraging the projections from DELPHI-pred.

We model ventilator pooling as a multi-period resource allocation over $S$ states and $D$ days. The model takes as input ventilator demand in state $s$ and day $d$, denoted as $v_{s,d}$, as well as parameters capturing the surge supply from the federal government and the extent of inter-state collaboration. We formulate an optimization problem that decides on the number of ventilators transferred from state $s$ to state $s'$ on day $d$, and on     the number of ventilators allocated from the federal government to state $s$ on day $d$. We propose a bi-objective formulation. The first objective is to minimize ventilator-day shortages; for robustness, we consider both projected shortages (based on demand forecasts) and worst-case shortages (including a buffer in the demand estimates). The second objective is to minimize inter-state transfers, to limit the operational and political costs of inter-state coordination. Mixed-integer optimization provides modeling flexibility to capture spatial-temporal dynamics and the trade-offs between these various objectives. We report the mathematical formulation of the model, along with the key assumptions, in the appendix.

\subsection{Results}

\begin{figure*}[h!]
    \centering
    \includegraphics[width=2\columnwidth]{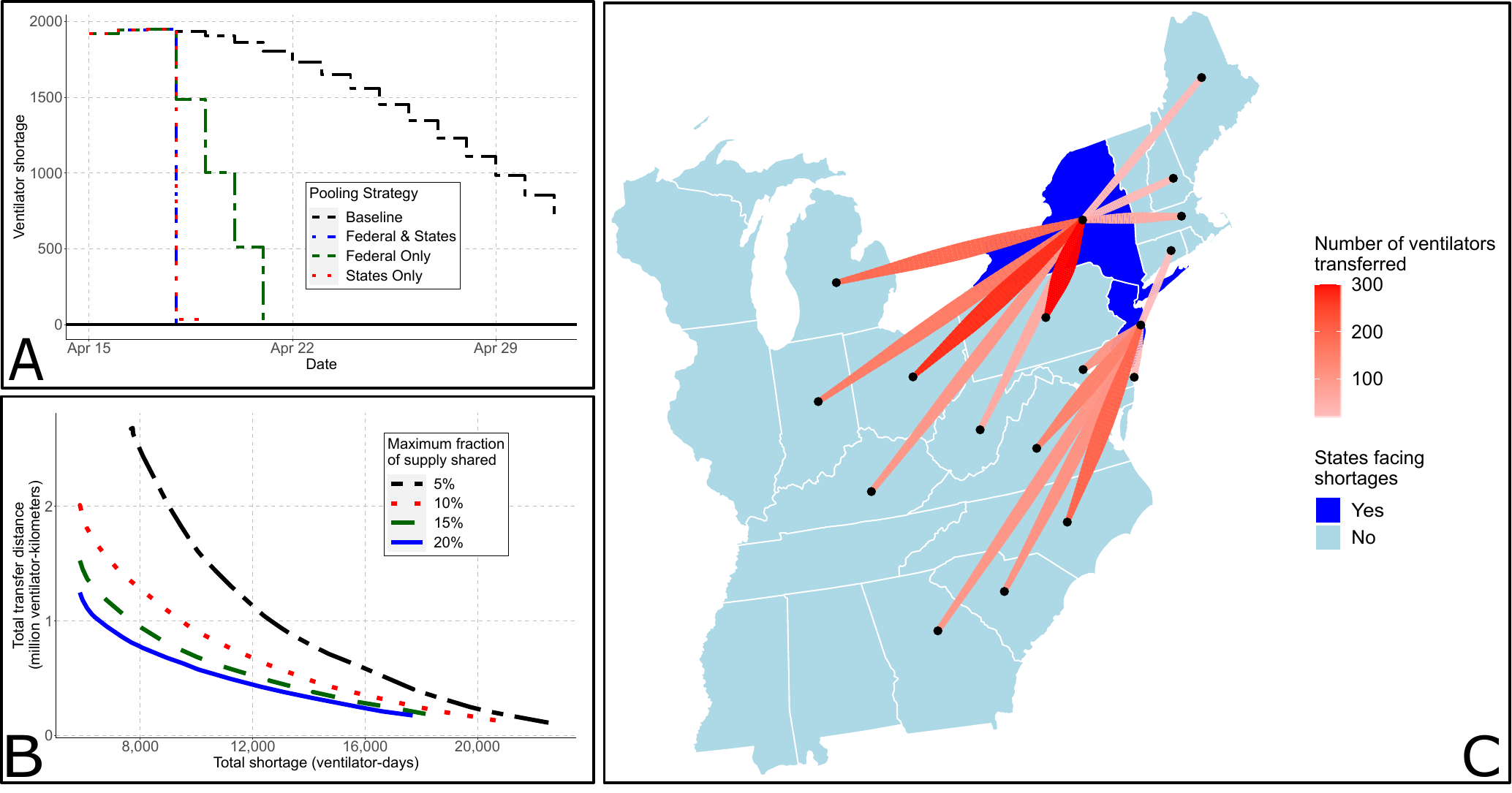}
    \caption{The edge of optimization to eliminate ventilator shortages. (A) Projected shortages (in ventilator-days) in a baseline setting (without transfers) and with optimized transfers between the states and/or from the federal government. (B) Pareto frontier between transfer distance and total shortage, for different state pooling fractions. (C) Map of inter-state transfers recommended on April 15 in the US Northeast. For clarity, we do not plot shortages of fewer than 5 ventilators and transfers of fewer than 10.}
    \label{fig:ventilator}
\end{figure*}

We implemented the model on April 15, a time of pressing ventilator need in the United States. We estimate the number of hospitalizations from DELPHI-pred as the sum of $DH_R$ and $DH_D$. From our clinical outcomes database in Section~\ref{sec:outcomes_database}, we estimate that 25\% of hospitalized patients are put on a ventilator, which we use to estimate the demand for ventilators. We also obtain the average length of stay from our clinical outcomes database (Figure~\ref{fig:demos_comborbs_vs_mort}).

Figure \ref{fig:ventilator}A shows the evolution of ventilator shortages with and without ventilator transfers from the federal government and inter-state transfers. These results indicate that ventilator pooling can rapidly eliminate all ventilator shortages. Figure \ref{fig:ventilator}C shows ventilator transfers recommended in the US Northeast on April 15 (with inter-state transfers only), overlaid on a map displaying the predicted shortage without transfers.

There are different pathways toward eliminating ventilator shortages. Figure \ref{fig:ventilator}B shows the trade-off between shortages and transfer distance---each line corresponds to the maximal fraction of its own ventilators that each state can pool. Overall, states do not have to share more than 10\% of their supply at any time to efficiently eliminate shortages. States can largely meet their needs with help from neighboring states, with cross-country transfers only used as a last resort. Broadly, results underscore trade-offs between ventilator shortages, the extent of inter-state transfers, the number of ventilators allocated from the federal government, and the robustness of the solution. We discuss these trade-offs further in the appendix.

\subsection{Discussion and Impact}

Our main insight is that ventilator shortages could be eliminated altogether through inter-state transfers and strategic management of the federal supply. Results also underscore (i) the benefits of inter-state coordination and (ii) the benefits of early coordination. First, ventilator shortages can be eliminated through inter-state transfers alone: leveraging a surge supply from the federal government is not required, though it may reduce inter-state transfers. Under our recommendation, the most pronounced transfers occur from states facing no shortages (Ohio, Pennsylvania, and North Carolina) to states facing strong shortages (New York, New Jersey). Second, most transfers occur in early stages of the pandemic. This underscores the benefits of leveraging a predictive model like DELPHI-pred to align the ventilator supply with demand projections as early as possible.

A similar model has been developed to support the re-distribution of ventilators across hospitals within the Hartford HealthCare system in Connecticut---using county-level forecasts of ventilator demand obtained from DELPHI-pred. This model has been used by a collection of hospitals in the United States to align ventilator supply with projected demand at a time where the pandemic was on the rise.

Looking ahead, the proposed model can support the allocation of critical resources in the next phases of the pandemic---spanning ventilators, medicines, personal protective equipment etc. Since epidemics do not peak in each state at the same time, states whose infection peak has already passed or lies weeks ahead can help other states facing immediate shortages at little costs to their constituents. Inter-state transfers of ventilators occurred in isolated fashion through April 2020; our model proposes an automated decision-making tool to support these decisions systematically. As our results show, proactive coordination and resource pooling can significantly reduce shortages---thus increasing the number of patients that can be treated without resorting to extreme clinical recourse with side effects (such as splitting ventilators).

\section{Conclusion}

This paper proposes a comprehensive data-driven approach to address several core challenges faced by healthcare providers and policy makers in the midst of the COVID-19 pandemic. We have gathered and aggregated data from hundreds of clinical studies, electronic health records, and census reports. We have developed descriptive, predictive and prescriptive models, combining methods from machine learning, epidemiology, and mixed-integer optimization. Results provide insights on the clinical aspects of the disease, on patients' infection and mortality risks, on the dynamics of the pandemic, and on the levers that policy makers and healthcare providers can use to alleviate its toll. The models developed in this paper also yield decision support tools that have been deployed on our dedicated website and that are actively being used by several hospitals, companies and policy makers.

\section*{Acknowledgments}
We would like to thank Dr. Barry Stein, Dr. Ajay Kumar, Dr.  Rocco Orlando, and Michelle Schneider from the Hartford HealthCare system, Dr. Angelo Pan, Dr. Rosario Canino, Sophie Testa and Federica Pezzetti from ASST Cremona, and HM Hospitals for discussions and data, as well as Hari Bandi, Katherine Bobroske, Martina Dal Bello, Mohammad Fazel-Zarandi, Alvaro Fernandez Galiana, Samuel Gilmour, Adam Kim, Zhen Lin, Liangyuan Na, Matthew Sobiesk, Yuchen Wang and Sophia Xing from our extended team for helpful discussions.

\bibstyle{pnas-new}
\bibliography{ref.bib, outcomes_db.bib}

\end{document}